\documentclass[12pt,preprint]{emulateapj}

\begin{document}

\title{Multiplicity at the Stellar/Substellar Boundary in Upper Scorpius}

\author{Adam L. Kraus (alk@astro.caltech.edu), Russel J. White (rjw@astro.caltech.edu), and Lynne A. Hillenbrand (lah@astro.caltech.edu)}
\affil{California Institute of Technology, Department of Astrophysics, MC 105-24, Pasadena, CA 91125}

\begin{abstract}

We present the results of a high-resolution imaging survey of 12 brown 
dwarfs and very low mass stars in the closest ($\sim$145 pc) young 
($\sim$5 Myr) OB association, Upper Scorpius.  We obtained images with the 
Advanced Camera for Surveys/High Resolution Camera on HST through the 
$F555W$ \,($V$), $F775W$ \,($i'$), and $F850LP$ \,($z'$) filters. This 
survey discovered three new binary systems, including one marginally 
resolved pair with a projected separation of only 4.9 AU, resulting in an 
observed binary fraction of $25\pm14\%$ at separations $\ga$4 AU. After 
correcting for detection biases assuming a uniform distribution of 
mass ratios for $m_s/m_p>0.6$, the estimated binary fraction is 
$33\pm17\%$. The binary fraction is consistent with that inferred for 
higher-mass stars in Upper Sco, but the separation and mass ratio 
distributions appear to be different. All three low-mass binary systems in 
Upper Sco are tight ($<18$ AU) and of similar mass ($m_s/m_p\ga0.6$), 
consistent with expectations based on previous multiplicity studies of 
brown dwarfs and very low mass stars in the field and in open clusters. 
The implication is that the distinct separation and mass ratio 
distributions of low-mass systems are set in the formation process or at 
very young ages, rather than by dynamical disruption of wide systems at 
ages $\ga$5 Myr. Finally, we combine the survey detection limits with 
the models of Burrows et al. (1997) to show that there are no planets or 
very low-mass brown dwarfs with masses $>10 M_J$ at projected separations 
$>20$ AU, or masses $>5 M_J$ at projected separations $>40$ AU orbiting 
any of the low-mass (0.04-0.10 $M_\sun$) objects in our sample. 

\end{abstract}


\section{Introduction}

Brown dwarfs (BDs) are objects with masses intermediate between those of 
stars and planets. As such, studies of BDs potentially can offer a unique 
link between theories of star formation and planet formation. Several BD 
formation scenarios have been proposed, but current observational results 
indicate that the properties of BDs, including their spatial distribution 
in star-forming regions \citep{luh04b, bric02}, velocity dispersion 
\citep{JG01,WB03}, accretion and inner disk frequency 
\citep{muen01,jay03,liu03,moh05}, and evidence for an embedded phase 
\citep{wh04}, are consistent with those of low-mass stars. Although this 
strongly supports the idea that BDs form in a fashion similar to stars, 
via the dynamical collapse and fragmentation of a cloud core, there is 
emerging evidence that the binary properties of brown dwarfs are 
fundamentally different from those of low-mass stars. 

Multiplicity surveys of field T dwarfs \citep{burg03}, L dwarfs 
\citep{K99,R01,C03,bouy03,G03}, late M dwarfs \citep{sieg05} and 
intermediate-age BDs in the Pleiades \citep{mart00} have found lower 
binary frequencies and smaller binary separations than for field G and 
early M dwarfs \citep{duq91,fisch92}. It is unclear if these binary 
properties reflect differences in the formation mechanism or the dynamical 
disruption of wide substellar binaries. One way to distinguish between 
these scenarios is by studying the properties of young BD binaries, 
possibly before any significant dynamical evolution has occurred. In this 
paper we present the results of an imaging multiplicity survey of brown 
dwarfs in the nearest ($\sim$145 pc) young ($\sim$5 Myr)\citep{pz99} OB 
Association, Upper Scorpius.

\section{Observations and Data Reduction}

Our targets were selected from the survey for low-mass members of Upper
Sco by \citet{amb00}.  They identified 15 candidate members with
spectral types of M5.5 or cooler, based on either low resolution spectra
(10 objects) or $R-I_C$ colors (5 objects).  Three of these 15 (USco 85,
114, 121) were determined to be likely non-members, based on low lithium
abundance and radial velocities that are inconsistent with higher mass
members \citep{muz03, white05}.  The remaining 12 were observed in
this program.  Subsequently, membership for all but the faintest member of
this sample, USco 137, has been confirmed with additional high resolution
spectra \citep{moh05}. USco 137 has not yet been observed 
spectroscopically. Since our targets include both very low mass stars and 
brown dwarfs, we hereafter refer to them as very low mass objects, or 
VLMOs.

Our images were obtained with the Advanced Camera for Surveys/High 
Resolution Camera on the Hubble Space Telescope, which has a field of 
view of 26x29\arcsec and distortion-corrected pixel size of 27 mas 
pix$^{-1}$. Observations were made between July and September 2003 with 
the filters F555W (V), F775W (i'), and F850LP (z') at two dither positions 
near the center of the detector and with two exposures per position. Total 
integration times were 510, 300, and 200 seconds, respectively. The F555W 
exposure times for the brightest objects (USco-55, 66, 67, and 75) was 
reduced to 350 seconds to allow for additional short exposures in F775W 
and F850LP, which were close to the saturation limit in the full-length 
exposures. We chose the V band to maximize angular resolution 
(diffraction limit $\theta_{res,V}=58$ mas) and the i' and z' bands to 
maximize sensitivity to very low mass companions. 

The raw images were calibrated and distortion-corrected by the CALACS 
pipeline during on-the-fly-reprocessing \citep{mack03}. Some cosmic rays 
remained, but their morphologies were substantially different from stellar 
PSFs, so they were easily identified by visual inspection. 

Potential point sources were identified with the IRAF task 
DAOPHOT/DAOFIND, which found all local brightness maxima with a 
significance of $>5\sigma$ and a full-width at half-maximum (FWHM) near 
the expected value for the filter. We then measured aperture photometry 
and point-spread function (PSF) fitting photometry for all objects in 
each field, and we report aperture photometry for all isolated objects and 
PSF photometry for all close binaries. PSF magnitudes were corrected to 
match aperture magnitudes based on results for the nine 
well-sampled objects that appeared isolated under visual 
inspection. Aperture photometry was carried out with the IRAF task 
DAOPHOT/PHOT with a 5-pixel aperture for faint objects and a 10-pixel 
aperture for bright objects, and our sky annulus had an inner radius of 
200 pixels and width of 15 pixels. We used the finite aperture corrections 
of \citet{sir04}. Point-spread function (PSF) fitting photometry was 
carried out with the IRAF task DAOPHOT/ALLSTAR \citep{stet87}. The PSF for 
each filter was constructed from the 9 isolated VLMOs. Since all of 
the targets were located near the center of the chip and have similar 
temperatures and extinctions, location and color effects should not be 
important. There was some variation in the PSF FWHM from target to target 
($\pm5\%$), which we attribute to small orbit-to-orbit changes in focus. 
However, the radial shape of the PSF was very similar for all objects 
except USco-109, which we will discuss in more detail in Section 3.1. 

Transformations to ground-based magnitudes ($V$, SDSS $i'$, and SDSS 
$z'$) were calculated with the IRAF task SYNPHOT/CALCPHOT, which 
convolves an input spectrum with transmission curves for HST's optics and 
filters or standard ground-based filters. Since SYNPHOT does not include 
transmission curves for the i' and z' filters, which are defined at the 
United States Naval Observatory 40-in telescope, they were obtained from 
the website for the Sloan Digital Sky 
Survey\footnote[1]{http://www.sdss.org/dr1/algorithms/standardstars/Filters/response.html} 
and represent the filters and optics at the USNO-40 observing at 1.3 
airmasses \citep{fukug96}. Based on transformations determined for a set 
of M0-M8 dwarfs and M5-M8 giants from the Bruzual-Persson-Gunn-Stryker 
Spectrophotometry Atlas \citep{bpgs96}, we find constant corrections 
that do not depend significantly on temperature or surface gravity: 
$m_{555}-m_V=-0.16\pm0.03$, $m_{775}-m_{i'}=+0.07\pm0.03$, and 
$m_{850}-m_{z'}=+0.03\pm0.03$. The uncertainties are estimated from the 
standard deviation between all tested objects. The transformed 
magnitudes\footnote[2]{The $V$ flux for USco-112 appears to be anomalously 
bright. Based on its $i'$ and $z'$ magnitudes, it appears to be of similar 
brightness and color to USco-55 B. However, it is 0.88 $\pm$ 0.06 
magnitudes brighter than USco-55 B in $V$. Exposures in both dither 
positions give consistent fluxes, so it is probably not the result of a 
cosmic ray hit on one exposure. We suggest that this is caused either by a 
transient optical brightening of the system (e.g. a flare or increased 
accretion) or by contamination from a spatially unresolved blue background 
object.} are listed in Table 1. The statistical uncertainties correspond 
to either the photon noise (for aperture photometry; typically $<$0.01 
magnitudes) or the goodness of fit (for PSF-fitting photometry; 0.02-0.09 
magnitudes). Systematic uncertainties in the magnitude 
transformations and aperture corrections are $\sim$0.03 magnitudes.

\begin{deluxetable*}{lllccccccc}
\tabletypesize{\scriptsize}
\tablewidth{0pt}
\tablecaption{Very Low Mass Objects in Upper Scorpius\label{tbl1}}
\tablehead{\colhead{Name}&\colhead{Date\tablenotemark{a}}&
\colhead{$V$\tablenotemark{b}}&\colhead{$i'$\tablenotemark{b}}&\colhead{$z'$\tablenotemark{b}}&\colhead{SpT\tablenotemark{c}}&\colhead{M 
(M$_\sun$)\tablenotemark{d}}}
\startdata
USco-55 A&49.9&18.73$\pm$0.04&15.92$\pm$0.012&14.81$\pm$0.09&M5.5&0.10$\pm$0.03\\
USco-55 B&49.9&18.92$\pm$0.015&16.19$\pm$0.08&15.12$\pm$0.09&(M6.0)&0.07$\pm$0.02\\
USco-66 A&62.6&18.92$\pm$0.021&16.36$\pm$0.016&15.41$\pm$0.012&M6.0&0.07$\pm$0.02\\
USco-66 B&62.6&18.94$\pm$0.022&16.30$\pm$0.023&15.29$\pm$0.008&(M6.0)&0.07$\pm$0.02\\
USco-67&49.0&18.47$\pm$0.002&15.52$\pm$0.001&14.32$\pm$0.001&M5.5&0.10$\pm$0.03\\
USco-75&59.7&18.71$\pm$0.002&15.80$\pm$0.001&14.61$\pm$0.001&M6.0&0.07$\pm$0.02\\
USco-100&67.0&19.09$\pm$0.002&15.99$\pm$0.001&14.73$\pm$0.001&M7.0&0.05$\pm$0.01\\
USco-109 A&64.6&20.20$\pm$0.03&17.13$\pm$0.04&15.87$\pm$0.016&M6.0&0.07$\pm$0.02\\
USco-109 B&64.6&21.21$\pm$0.076&18.07$\pm$0.08&16.92$\pm$0.04&(M7.5)&0.04$\pm$0.01\\
USco-112&71.6&18.04$\pm$0.001\tablenotemark{e}&16.15$\pm$0.001&15.17$\pm$0.001&M5.5&0.10$\pm$0.03\\
USco-128&70.6&21.29$\pm$0.007&17.79$\pm$0.002&16.34$\pm$0.002&M7.0&0.05$\pm$0.01\\
USco-130&87.8&21.39$\pm$0.008&17.80$\pm$0.002&16.30$\pm$0.002&M7.5\tablenotemark{f}&0.04$\pm$0.01\\
USco-131&63.6&21.98$\pm$0.011&18.22$\pm$0.002&16.67$\pm$0.002&M6.5\tablenotemark{f}&0.06$\pm$0.01\\
USco-132&88.8&22.11$\pm$0.013&18.31$\pm$0.002&16.59$\pm$0.002&M7.0&0.05$\pm$0.01\\
USco-137&88.9&22.92$\pm$0.023&19.27$\pm$0.004&17.75$\pm$0.003&M7.0\tablenotemark{f}&0.05$\pm$0.01\\
\enddata
\tablenotetext{a}{Observation Date: JD minus 2452800.}
\tablenotetext{b}{Uncertainties are statistical; systematic uncertainties due to aperture corrections and conversion to standard systems are $\sim$0.03 magnitudes.}
\tablenotetext{c}{Spectral types in parentheses are inferred from photometry presented here; others are from Ardilla et al.(2000) (Section 4.1)}
\tablenotetext{d}{Masses are determined from the models of Baraffe et al. (1998).}
\tablenotetext{e}{The V magnitude for USco-112 is anomalously bright. (Section 2)}
\tablenotetext{f}{Spectral type is based only on photometric $R-I_C$ colors.}
\end{deluxetable*}

\section{Results}

\subsection{New VLMO Binaries}

In Figure 1, we present contour plots of three candidate binaries 
(USco-55, 66, and 109) and the apparently single star USco-67 in the 
F555W, F775W, and F850LP filters. The USco-55 and USco-66 systems are 
clearly resolved. The USco-109 system is not obviously resolved and was 
not initially reported as a double-source by ALLSTAR, but the PSF appears 
to be marginally elongated in the +x direction relative to the 
single VLMO USco-67. 

One limitation in the ALLSTAR-based data reduction method is that binaries 
with very close ($\la$$\lambda$$/D$) separations are often not identified, 
even when their combined PSF is elongated at a high confidence level. 
DAOFIND, the task which identifies potential objects in the images, only 
identifies point sources based on the presence of a distinct peak. Thus, 
automated photometry will be biased against the detection of very close 
binaries. This limitation can be overcome for known or suspected 
binaries by manually adding a second point source in approximately the 
correct location and letting ALLSTAR recenter it to optimize the fit; if 
it does not produce a statistically significant fit, ALLSTAR then discards 
it. We have done this for USco-109.

In Figure 2, we present plots for USco-109 and the next-brightest object 
in the same images, a likely background field star (Section 3.2). The 
first three columns show USco-109 and the residuals from fitting with one 
and then two point sources, and the last two columns show the field star 
and its single-source fit. The maximum and minimum pixel values are also 
given to allow quantitative comparison of the residuals to the original 
images. The common position angle of the residuals in the single-source 
fit in all three filters seems to imply that this elongation is a real 
effect, and not simply noise. USco-109 appears elongated in the same 
direction relative to both its neighbor and USco-67; this extension is 
therefore unlikely to be an artifact due to excess jitter, which would 
affect the field star as well. USco-109 is better fit with two 
point sources in all three filters, and the fit reduces the residuals by 
factors of 2-8. Moreover, these fits independently find similar positions 
and flux ratios in all three filters, which further supports its 
classification as a binary system. Since the similar two-source residuals 
in each filter imply some remaining uncertainty in the fit and the 
separation is only $\sim$1.3 pixels, followup observation of this system 
to confirm its multiplicity and properties should be a priority. However, 
we will proceed under the assumption it is a binary system in our 
subsequent analysis. 

We summarize the properties of these three systems in Table 2. The 
uncertainties in separation and position angle are determined from the 
standard deviations in the locations as reported by ALLSTAR for the three 
filters.

\begin{deluxetable*}{ccccccc}
\tabletypesize{\scriptsize}
\tablewidth{0pt}
\tablecaption{Binary Parameters\label{tbl2}}
\tablehead{\colhead{Name} & \colhead{Projected} & \colhead{Projected} & 
\colhead{Position} & \colhead{$\Delta$$SpT$} & \colhead{$q$}
\\
\colhead{} & \colhead{Separation (mas)} & \colhead{Separation(AU)} & 
\colhead{Angle(deg)}
}
\startdata
USco-55 
AB&121.6$\pm$0.6&17.63$\pm$0.09&307.7$\pm$0.4&0.5$\pm$0.1&0.86$\pm$0.04\\
USco-66 AB&70.3$\pm$0.5&10.19$\pm$0.07&31.7$\pm$0.2&0.0$\pm$0.1&0.94$\pm$0.03\\
USco-109 AB&34$\pm$2&4.9$\pm$0.3&302$\pm$3&1.5$\pm$0.2&0.59$\pm$0.04\\
\enddata
\end{deluxetable*}

\subsection{VLMOs and Background Stars}

In Figure 3, we present an $i'$ versus $i'-z'$ color magnitude diagram of 
the 12 primary targets (filled circles) and all other objects detected at 
the $5\sigma$ level in both filters. Objects within 5\arcsec (725 AU), 
which are statistically more likely to be physically associated, are shown 
as open circles while other objects are shown as crosses. Also shown are 
the average main sequence at the distance of Upper Sco \citep{hawl02} and 
a 5 Myr isochrone based on the evolutionary models of \citet{bcah98}. The 
location of the isochrone is determined by converting the predicted $I_C$ 
and $J$ magnitudes to $i'$ and $z'$ magnitudes, using $I_C-i'$ colors 
derived using the methods described in Section 2 and $z'-J$ colors found 
from the SDSS field main sequence for older, more massive M dwarfs by 
\citet{hawl02}.  Although the $I_C-i'$ transformations should be accurate 
(Section 2), the $z'-J$ transformations may be more sensitive to surface 
gravity differences due to the larger difference in central wavelengths. 
Consequently, the $i'-z'$ color of the 5 Myr isochrone is somewhat 
uncertain.

The three very close companions to USco-55, USco-66, and USco-109 are 
located above the SDSS main sequence and well above the background 
population, so they are most likely association members. Based on their 
close proximity and the low density of association members, we conclude 
that these are physically associated companions. All other objects fall 
well below the empirical main sequence and are most likely background 
stars. The marginally resolved companion to USco-109 resides near the 
empirical (but somewhat uncertain) main sequence, but given the 
uncertainty in its $i'-z'$ color, its location is consistent with the 
broad luminosity distribution of low-mass Upper Sco members that we 
discuss in Section 5.

\subsection{Sensitivity Limits}

We determined detection limits as a function of distance from the primary 
stars via a Monte Carlo simulation similar to that of \citet{met03}. We 
used the IRAF task DAOPHOT/ADDSTAR and our average PSF to add artificial 
stars at a range of radial separations and magnitudes to the fields of 
USco-67, USco-128, and USco-137, which represent the maximum, median, and 
minimum brightness sources in our sample. We then attempted to identify 
them with our ALLSTAR 
photometry pipeline. In Figure 4, we show the detection limits 
for the USco-67 and USco-128 fields in V and z', as a function of 
separation, at which we can detect $>10\%$, $50\%$, and $90\%$ of the 
companions. At small separations ($\la$20 AU), the 50\% detection 
thresholds roughly scale with the brightness of the primary; they are 
similar for all objects in terms of $\Delta$$m$. The detection limits 
converge to constant values at large separations where the background 
dominates the noise: $z'$=23.40$\pm$0.05, $i'=$25.08$\pm$0.05, and 
$V$=26.01$\pm$0.05 (uncertainties are based on the standard deviation 
between the three fields). The simulations demonstrate that we potentially
could identify bright pairs as close at 1 pixel (0.027\arcsec; 4 AU) 
but none closer. We also show the locations (in $\Delta$$m$ and 
separation) of the three companions to USco-55, 66, and 109. These results 
indicate that the probability of detecting a system with the separation 
and flux ratio of USco-109 with automated photometry is between 10\% and 
50\% for the brightest objects and becomes negligible for the fainter 
targets in our sample. However, as discussed in Section 2, these limits 
are conservative at small separations since ALLSTAR often reports only a 
single point source when the PSF is still noticeably elongated and manual 
fitting can distinguish two point sources. The position angle of the 
binary can also play a role in the detection probability; as can be seen 
in Figure 1, the blended PSF of USco-109 is elongated along the axis where 
the PSF is narrowest, providing the most significant possible detection. 
Artificial star tests show that if the companion was near one of the 
extended lobes of the PSF, it would have been much more difficult to 
identify.

\subsection{Uncertainties in Binary Properties}

A similar Monte Carlo routine was used to test the uncertainties 
in the measurements of USco-109. We used ADDSTAR and the average PSF to 
construct 100 simulated images, given the positions and brightnesses 
reported for the primary and the secondary for the real images, and then 
used ALLSTAR to perform PSF-fitting photometry on these simulated images. 
The standard deviation in separation ($\sim$1 mas) from the simulated 
images is consistent with that calculated from the standard deviation in 
separation between the three filters ($\sim$2 mas), and the standard 
deviation in the flux ratio $\Delta$$m$ (0.09 magnitudes in $V$, 0.06 in 
$i'$ and $z'$) is consistent with that determined from the 
magnitudes reported by ALLSTAR for each object. 

\section{Analysis}

\subsection{Inferred Properties}

In Table 1, we give the inferred spectral types and masses for all of the 
VLMOs in our sample. Spectral types for single VLMOs and for binary 
primaries in our sample are taken from \citet{amb00}. Spectral types for 
nine objects were determined from low-resolution spectra; the other three 
(USco-130, 131, and 137) have only photometric spectral types as 
determined from $R-I_C$ colors.  Ardila et al. found that although this 
region is young, the extinction is generally low ($A_V<1$), so the 
photometric spectral types should be reasonably accurate. Spectroscopic 
observations of all targets except USco-137 by \citet{moh05} appear to be 
consistent with these assigned spectral types.

The masses for this sample are estimated from the 5-Myr 
mass-magnitude-temperature relations of \citet{bcah98} and the 
temperature-SpT relations of \citet{luh03}, and range from 0.04 to 0.10 
$M_\sun$. Large systematic errors may be present in these and all pre-main 
sequence models (e.g. Baraffe et al. 2002; Close et al. 2005; Reiners et 
al. 2005), so they are 
best used for relative comparison only. The models of Baraffe et al. 
predict that the mass ratio ($q=m_{s}/m_{p}$) and $\Delta$$SpT$ for the 
three binaries are a function of the primary-to-secondary flux ratios 
$\Delta$$m$ with only a minor mass dependence; we report these quantities 
as determined from the flux ratio $\Delta$$i'$ in Table 1. The 
uncertainties reflect the uncertainties in the flux ratios, but do not 
include any systematic uncertainties from the models. The values 
determined from $\Delta$$V$ and $\Delta$$z'$ are consistent with these 
results, but the values for $\Delta$$i'$ are least dependent on primary 
mass, so they are most reliable.

\subsection{The Binary Frequency at the Stellar/Substellar Boundary}

For our 12 M5.5-M7.5 targets, we have detected three binary VLMO systems 
with projected separations of 4.9-17.6 AU, giving an observed binary 
fraction of 25$\pm$14$\%$. However, as can be seen in Figure 4, the 
detection of faint companions is difficult at separations comparable to 
the PSF width (58 mas in V). Consequently, the total binary 
fraction may be higher. We can estimate the number of undetected 
companions via a method based on that of \citet{C03} where we convolve an 
assumed mass ratio distribution with the mass-magnitude relations of 
\citet{bcah98} and the detection thresholds $\Delta$$m$ as calculated in 
section 4.3. Similar results are found for all three filters; the 
following results are those for the $i'$ band.

Observations of field M-dwarfs have found a flat mass ratio 
distribution from 1.0 to 0.1 \citep{fisch92} for all companions 
at separations $a>4$ AU. However, studies of field VLMOs have found only 
BD companions with mass ratios $q>0.6$ ($\Delta$$m<1$). If we adopt a mass 
ratio distribution that is uniform between 0.6 and 1.0 and zero for 
$q<0.6$, then the detection thresholds in Figure 4 can be used to predict 
the number of unidentified companions. The $50\%$ detection limit spans 
the full range of possible mass ratios at separations $>$10 AU, and then 
approaches $\Delta$$m=0$ at a separation of $\sim$4 AU. The average 
threshold of $\Delta$$m=0.5$ in the separation range $4<a<10$ AU 
corresponds to a detection threshold of $q=0.8$. This implies that all 
companions outside 10 AU and $\sim$1/2 of all companions inside 10 AU were 
detected. Since we detected one companion within 10 AU, we predict the 
existence of one additional companion, which yields a 
completeness-corrected binary fraction for separations $\ga$4 AU of 
$33\pm17\%$. If we assume a flat mass ratio distribution between 0.1 and 
1.0, as for field stars, then similar analysis finds 1/2 
of an undetected companion for separations $10<a<15$ AU and 2 companions 
for $a\la10$ AU, for a complete binary fraction of $46\pm25\%$. However, 
the paucity of observed unequal-mass companions strongly implies that the 
lower value is more appropriate, and even it might be an overestimate 
since our Monte Carlo tests produced conservative limits at small 
separations and USco-109 lies below our nominal detection curve.

We note that the intrinsically higher luminosity of binaries makes them 
easier to identify in membership surveys, which can bias the binary 
frequency to larger values. However, this should not significantly bias 
our results since all targets except USco-137 are substantially brighter 
than the detection limit of the original survey by Ardila et al.

\subsection{Limits on Planetary-Mass Companions}

The high dynamic range observations of young stars and brown dwarfs 
presented here have the potential to directly image wide planetary-mass 
companions. In Figure 4, we indicate the predicted zero-extinction 
brightness of some representative masses of planetary companions based on 
the models of \citet{bur97}; since most of our targets have $A_V<1$, 
extinction corrections at i' and z' should be small ($<0.5$ magnitudes). 
The extremely red colors predicted for planetary-mass objects imply that 
z' observations provide the strictest limits on planetary companions. 
Based on the lack of detections, we conclude that there are no planetary 
companions with mass $\ga$5 $M_J$ at projected separations larger than 280 
mas (40 AU) or mass $\ga$10 $M_J$ at projected separations larger than 140 
mas (20 AU) among our sample. For comparison, these limits would have 
allowed for marginal detection of the candidate planetary companion to 
2MASSWJ 1207334-393254, a substellar member of the TW Hya association, 
which has a projected separation of $\sim$54 AU, a flux ratio of 
$\Delta$$z'$$\sim7$, and a predicted mass of $\sim$5 $M_J$ 
\citep{chauv04}.

\section{Discussion}

Multiplicity surveys of VLMOs have suggested a fairly uniform set 
of binary properties in the field \citep{C03,bouy03,burg03,sieg05} and in 
the Pleiades \citep{mart03}. No companions with wide separations ($>$20 
AU) or with unequal mass ratios ($q<0.7$) were found, despite sufficient 
sensitivity for their detection. The binary fractions observed were also 
significantly lower than the binary fractions of 57\% for field G-dwarfs 
\citep{duq91} and 35-43\% for field M-dwarfs \citep{reid97,fisch92}. 
Specifically, Close et al. found a binary fraction of $15\pm7\%$ for 
separations of $\ga$3 AU for field late-M and L dwarfs, Bouy et al. found 
a binary fraction of $18.8\pm3.7\%$ at separations of $\ga$1 AU for field 
L dwarfs, Burgasser et al. found a binary fraction of $9^{+15}_{-5}\%$ at 
separations of $\ga$1 AU for field T dwarfs, Siegler et al. found a 
binary fraction of $9^{+4}_{-3}\%$ for field M6-M7.5 dwarfs at 
separations of $\ga$3 AU, and Martin et al. found a binary fraction of 
$15^{+15}_{-5}\%$ for separations of $\ga$7 AU for Pleiades BDs. 

The study presented here is the first survey of young BDs with the 
resolution and sensitivity to identify binaries as close as 4 AU, allowing 
a more direct comparison with field surveys. Our method of correcting for 
detection biases is based on that of \citet{C03}, so our 
completeness-corrected binary fraction can be compared most directly to 
that study. Our binary fraction, $33\pm17\%$ for separations $\ga$4 AU, is 
higher than, but statistically consistent with their 
completeness-corrected binary fraction for field VLMOs ($15\pm7\%$). 
However, the field surveys are sensitive to companions with separations as 
small as $\sim1$ AU and found many companions within 4 AU that we could 
not have detected because of the larger distance to our targets. Our 
binary fraction for companions with $a\ga4$ AU only sets a lower limit on 
the total fraction of VLMO binaries. Higher-resolution observations or 
long-term spectroscopic monitoring of these young VLMOs is needed to 
determine if they too have an abundance of even closer companions.

All three Upper Sco VLMO binaries have flux ratios of 
$\Delta$$m$$\la$$1.0$, corresponding to mass ratios of $\ga0.6$, which 
also agrees with with the other surveys described above, which found no BD 
binaries with $q<0.6$. The implication is that the mass ratio distribution 
is not flat at low masses, but instead is biased toward binaries of 
nearly-equal mass. This indicates that a completeness correction which 
assumes a flat mass ratio for q$\ga$0.1 would strongly overestimate the 
VLMO binary frequency. Like the field and open cluster surveys, we found 
no wide binary companions. Our results are also consistent with a survey 
of the young ($\sim$3 Myr) star-forming region IC 348 by \citet{luh04b}, 
which found a $1\sigma$ upper limit on the binary fraction of $\sim$7\% 
for separations $>$20 AU. Wide ($>20$ AU) companions are therefore much 
less common around VLMOs than around G and early M dwarfs, though some 
candidate wide companions have been identified in Chamaeleon I 
\citep{neu02,luh04a} and TW Hydra \citep{chauv04}.

The distinct binary properties of field BDs, relative to those of stars, 
indicate that these properties are mass dependent. However, the form of 
this dependence is not known; \citet{krou03} and \citet{C03} interpret 
current data as a break near the stellar/substellar boundary while 
\citet{luh04b} argues for a smooth mass dependence. We investigate the 
mass dependence at 
a young age by combining our results with a binary census of more massive 
T Tauri members of Upper Sco by \citet{koh00}. This survey targetted 118 
X-ray selected pre-main-sequence stars ranging from G5 to M5, 
corresponding to masses of $2.0 - 0.13 M_\sun$ and found a binary fraction 
of $35.2\pm6.3\%$ for companions with separations between $6\arcsec$ (870 
AU) and $0.13\arcsec$ (19 AU). Dividing the sample into spectral type bins 
reveals no spectral type dependence of the binary frequency, though the 
subsample sizes are not sufficient to completely rule out a gradual 
decline. Specifically, the survey found a binary frequency of 26/50 = 
$52\pm10\%$ for Upper Sco members of spectral type M0-M5 ($0.7-0.13 
M_\sun$) for separations between 19 and 870 AU, including several very 
wide companions to primaries in the M4-M5 range. Kohler et al. did not 
confirm association membership of the companions with colors or proper 
motions, but they estimate from the low background object density 
($6.64\pm0.45$ $deg^{-2}$ in K, comparable to our background object 
density of $7.8\pm0.8$ $deg^{-2}$ in i' and z') that no more than 15\% of 
their detected binaries were spurious claims. They also found no decline 
in mean or maximum separation and no preference toward equal-mass 
companions with decreasing mass, though the mass ratio result could not be 
tested for the lowest-mass targets since unequal-mass companions would 
have fallen below the detection limit. Our measurements in the mass range 
$0.04-0.10 M_\sun$ suggest a significant decrease in the number of wide 
binaries. Since none of the newly-discovered binaries presented here would 
have been found in the Kohler survey, there may be a discontinuity in the 
separation distribution at a mass of $\sim$0.10$M_{\sun}$. 

We also note that resolving the newly-discovered binaries does not reduce 
the large scatter in the color-magnitude diagram at low masses that has 
been observed in previous surveys \citep{amb00} and is evident for the 
apparently single VLMOs in Figure 1. \citet{pz99} argue for a consistent 
age of 5 Myr based on the narrow locus seen on color-magnitude diagrams of 
more massive Upper Sco members, but the poor fit between our data and any 
single isochrone of \citet{bcah98} indicates that the lowest-mass members 
of Upper Sco could have a significant range in ages or distances. The 
uncertainty introduced by transforming of the 5 Myr isochrone into the 
SDSS magnitude system makes it difficult to interpret other possibilities, 
such as the existence of additional unresolved binaries or preferential 
selection of older or younger systems in the original survey by Ardila et 
al.

In summary, the binary properties of the young VLMOs observed here are the 
first convincing evidence that the small separations and equal mass ratios 
of VLMO binaries are established at young ages rather than as the result 
of subsequent dynamical evolution after the T Tauri phase of evolution. 
The uncertainty in our total binary fraction is too large to directly test 
the suggestion that the binary fraction declines with declining 
mass, but if the total binary fraction of the field population is truly 
primordial, as the small separations and equal masses appear to be, then 
this provides powerful evidence in support of this assertion. However, 
complementary spectroscopic surveys are needed to determine if the 
apparent deficit of substellar binaries is a true absence or is due to a 
shift to smaller separations than can be detected with imaging alone. 
Finally, the recent discovery of candidate wide binary systems in the 
low-density star-forming regions Chamaeleon I \citep{neu02,luh04a} and TW 
Hydra \citep{chauv04} indicate that there could be differences in the VLMO 
binary formation process and initial conditions in OB associations 
compared to low-density T associations.

\acknowledgements

We would like to thank Andrea Ghez for helpful comments, and we thank the 
anonymous referee for a thorough and helpful response which improved the 
quality of this work. This work is based on observations made with the 
NASA/ESA Hubble Space Telescope, obtained at the Space Telescope Science 
Institute, which is operated by the Association of Universities for 
Research in Astronomy, Inc., under NASA contract NAS 5-26555. These 
observations are associated with program \#9853.

\begin{figure}
\epsscale{1.00}
\plotone{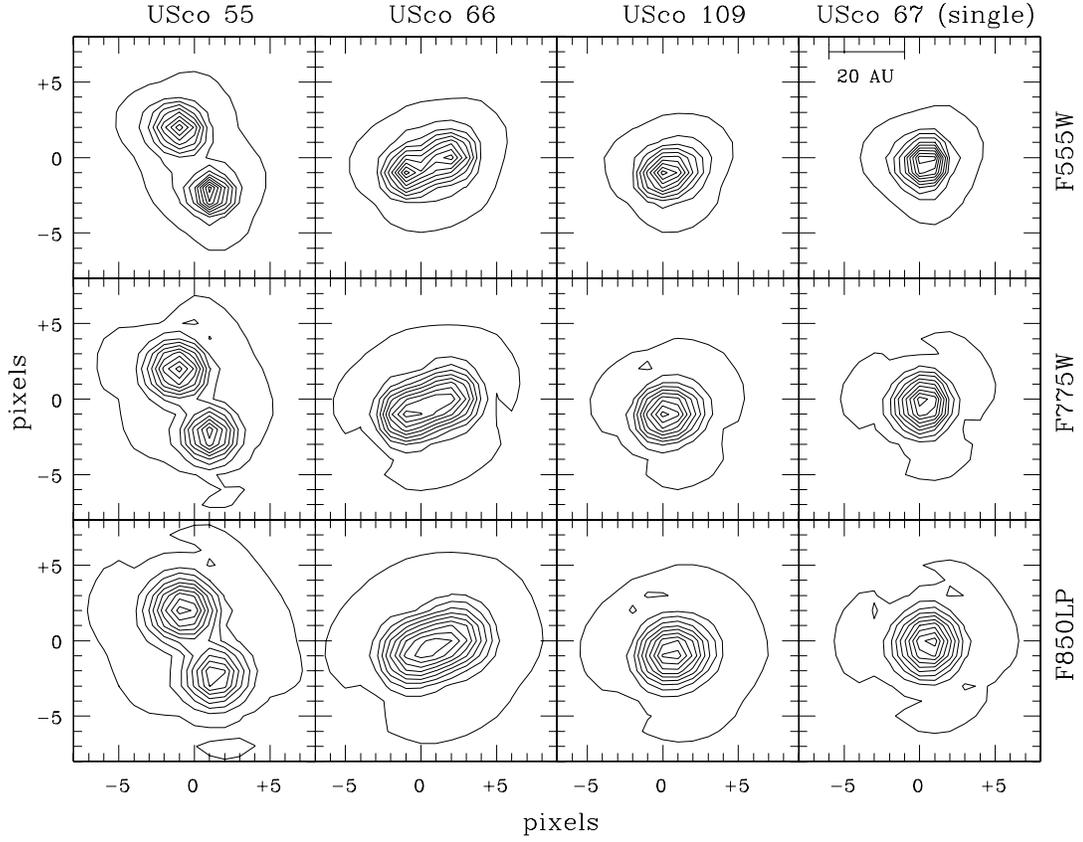}
\caption{
Contour plots of four targets: USco-55, USco-66, USco-109, and USco-67, 
respectively. Units are in pixels, and the projected physical scale at the 
distance of Upper Sco is shown in the upper right panel. Contours are 
drawn at 95\% through 5\% of the maximum pixel value, in increments of 
10\%. The field of view in each image is 432 mas, or $\sim$60 AU at 145 
pc.}
\end{figure}

\begin{figure}
\epsscale{1.00}
\plotone{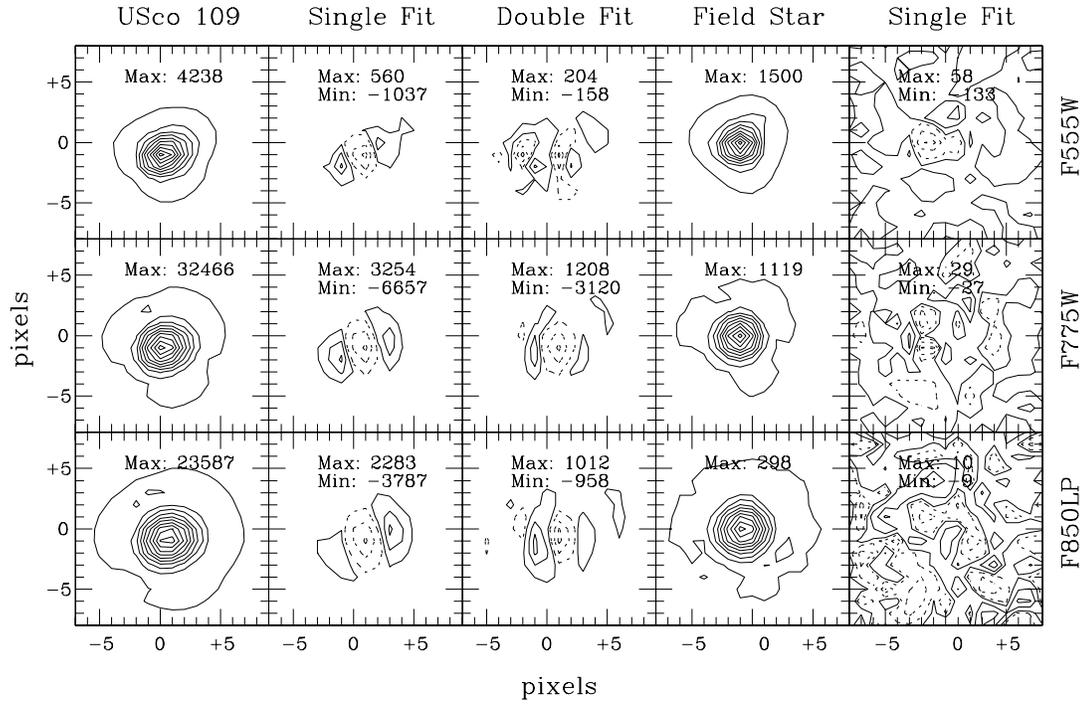}
\caption{
Contour plots of USco-109 and a background star in the same field for all 
three filters. The first three columns show USco-109 and the residuals 
from fitting with one and then two point sources, and the last two columns 
show the background star and the residuals from a single-source fit. For 
residuals, contours are drawn at the 90\%, 50\%, and 10\% levels of 
maximum (solid lines) and minimum (dashed lines). The maximum and minimum 
pixel values are given to allow comparison of the residuals to the 
original images. The pixel values in the last column, where the sky 
background contours fill each panel and obscure the text, are (58,-133), 
(29,-27), and (10,-9), respectively. 
}
\end{figure}

\begin{figure}
\epsscale{1.00}
\plotone{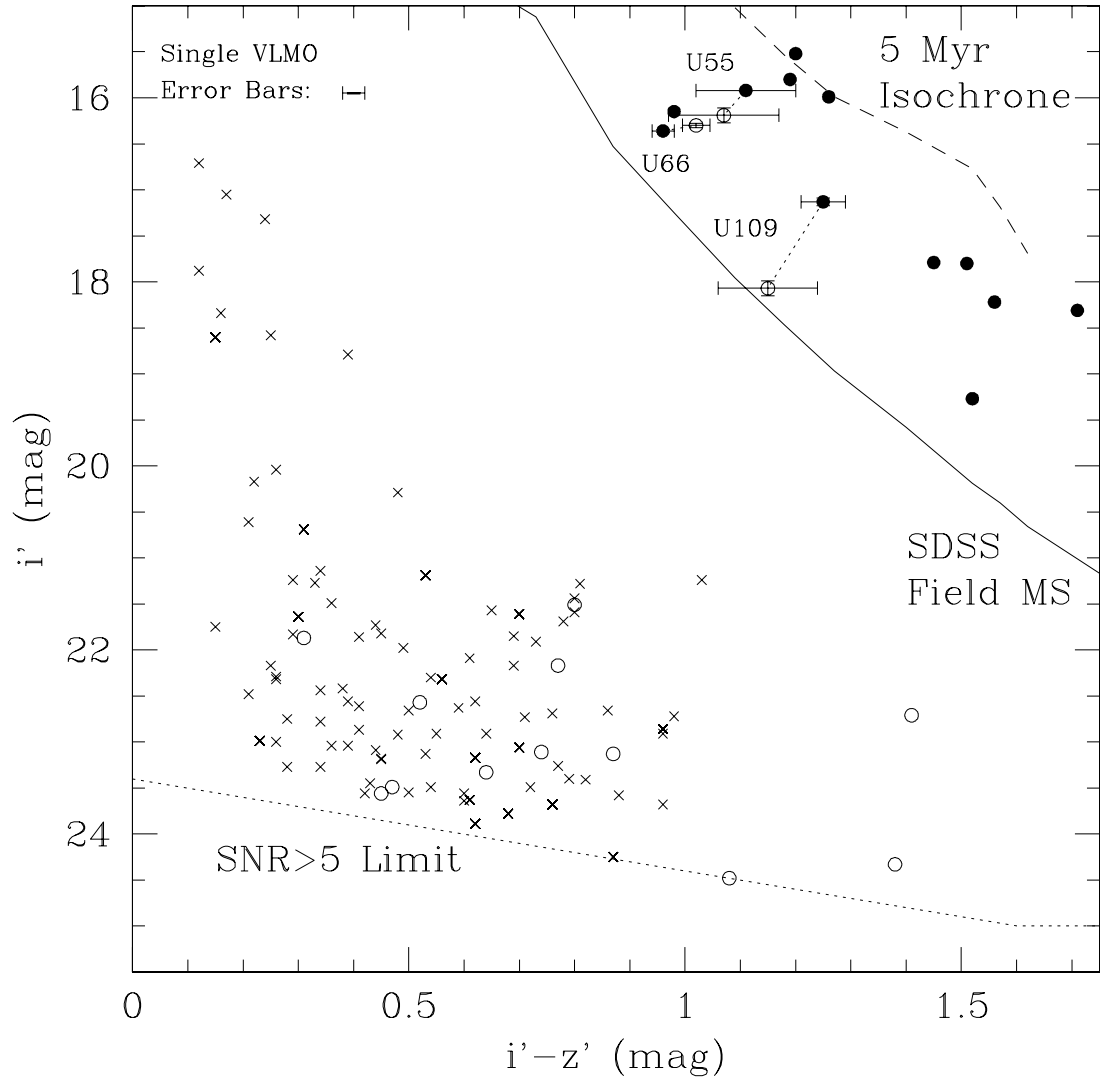}
\caption{An $i'$ vs $i'-z'$ color magnitude diagram. Primary Upper Sco 
targets are shown as filled circles, neighbors within 5\arcsec \,are shown 
as open circles, and widely-separated objects are shown as crosses. The 
SDSS field main sequence (solid line), a 5 Myr isochrone (dashed line), 
and the detection limits of the survey (dotted line) are also shown. 
Candidate binary VLMO pairs are connected with dotted lines and labeled. 
The error bars for binary components are associated with each point, and 
the error bars for single objects are shown in the upper left corner.
}
\end{figure}

\begin{figure}
\epsscale{1.00}
\plotone{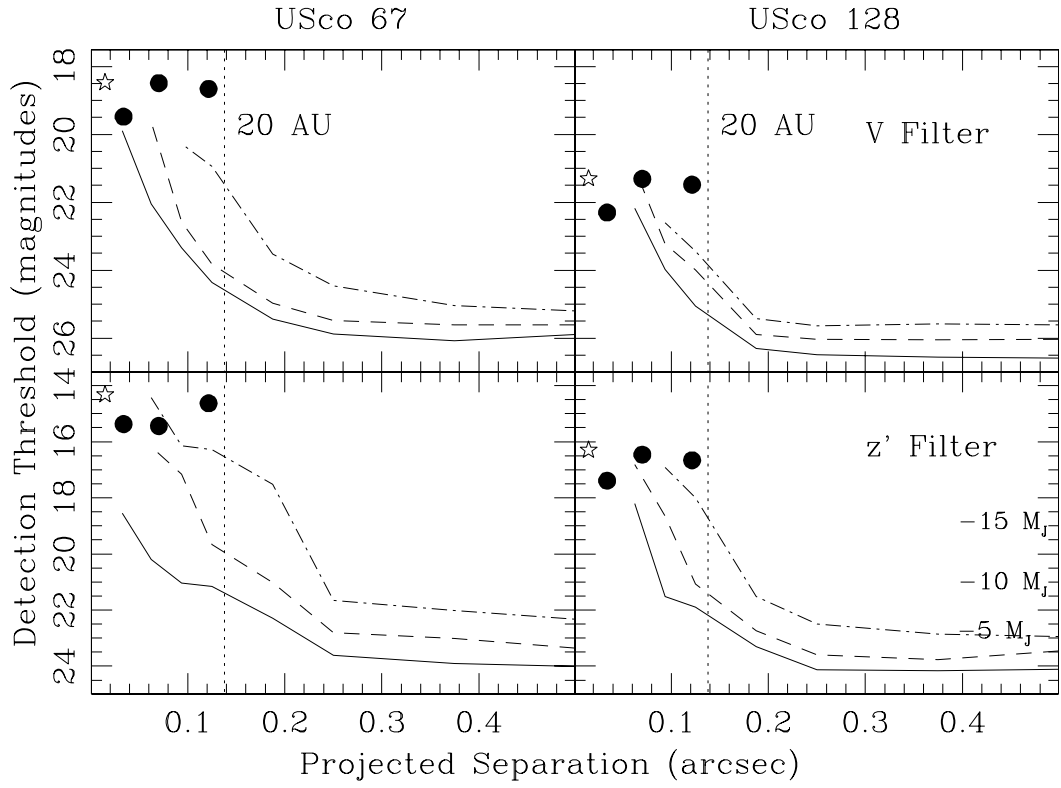}
\caption{Five-sigma detection frequencies (10\%: solid, 50\%: dashed, and 
90\%: dash-dotted) as a function of separation in each filter for the 
VLMOs of maximum and median brightness, USco-67 and USco-128. 
Corresponding brightnesses of potential planetary-mass companions are 
shown on the right for the z' plot. The brightness of the primary object 
is denoted with a star to allow conversion to $\Delta$$m$ values, and the 
vertical dotted line indicates a separation of 20 AU at the distance of 
Upper Sco (145 pc). The filled circles mark the separation and $\Delta$$m$ 
values for the three binary systems.
}
\end{figure}
\end{document}